\documentclass[pdflatex,sn-mathphys-num]{sn-jnl}
\usepackage{graphicx}%
\usepackage{array,multirow}%
\usepackage{amsmath,amssymb,amsfonts}%
\usepackage{amsthm}%
\usepackage{mathrsfs}%
\usepackage[title]{appendix}%
\usepackage{xcolor}%
\usepackage{textcomp}%
\usepackage{manyfoot}%
\usepackage{booktabs}%
\usepackage{algorithm}%
\usepackage{algorithmicx}%
\usepackage{algpseudocode}%
\usepackage{listings}%
\usepackage{geometry}
\begin{document}
%\title[title]{1 GBaud analog end-to-end in-situ adaptive training with locally-updated physical gradients on integrated photonic platform}
%\title[title]{1~TOPS/W analog end-to-end real-time online training with adaptive physical gradients on integrated photonic platform}

%\title[title]{1~TOPS/W analog end-to-end real-time training with online adaptive tracking on integrated photonic platform}

\title[title]{Fully analog end-to-end online training with real-time adaptibility on integrated photonic platform}

% Nature computational science (Jie Sun)
% Nature machine intelligence
% Nature electronics
% Nature Communications

\author*[1]{\fnm{Zhimu} \sur{Guo}}\email{15zg11@queensu.ca}

\author[1,3]{\fnm{A.} \sur{Aadhi}}

\author[2]{\fnm{Adam N.} \sur{McCaughan}}

\author[3]{\fnm{Alexander N.} \sur{Tait}}

\author[4]{\fnm{Nathan} \sur{Youngblood}}

\author[2]{\fnm{Sonia M.} \sur{Buckley}}

\author*[1,3]{\fnm{Bhavin J.} \sur{Shastri}}\email{shastri@ieee.org}

\affil*[1]{\orgdiv{Centre for Nanophotonics, Department of Physics, Engineering Physics, and Astronomy}, \orgname{Queen's University}, \orgaddress{\city{Kingston}, \postcode{K7L 3N6}, \state{ON}, \country{Canada}}}

\affil[2]{\orgname{National Institute of Standards and Technology}, \orgaddress{\city{Boulder}, \postcode{80305}, \state{CO}, \country{USA}}}

\affil[3]{\orgdiv{Smith Engineering, Department of Electrical and Computer Engineering}, \orgname{Queen's University}, \orgaddress{\city{Kingston}, \postcode{K7L 3N6}, \state{ON}, \country{Canada}}}

\affil[4]{\orgdiv{Department of Electrical and Computer Engineering}, \orgname{University of Pittsburgh}, \orgaddress{\city{Pittsburgh}, \postcode{15261}, \state{PA}, \country{USA}}}

\abstract{
Neuromorphic hardware has emerged as a transformative platform for artificial intelligence and machine learning by offering unprecedented speed and energy efficiency. Particularly, analog neuromorphic photonic processors are uniquely positioned to harness the ultrafast bandwidth and inherent parallelism of light, enabling scalability, on-chip integration and significant improvement in computational performance. However, major challenges remain unresolved in state-of-the-art architectures, especially in achieving real-time online training, efficient end-to-end anolog systems, and adaptive learning for dynamical environmental changes. Here, we demonstrate an on-chip photonic analog end-to-end adaptive learning system realized on a foundry-manufactured silicon photonic integrated circuit. Our platform leverages a multiplexed gradient descent (MGD) algorithm to perform in-situ, on-the-fly training, while maintaining robustness in online tracking and real-time adaptation---a critical step toward a practical neuromorphic hardware platform. At its core, the processor features a monolithic integration of a microring resonator weight bank array and on-chip photodetectors, enabling direct optical measurement of gradient signals. This eliminates the need for high-precision digital matrix multiplications, significantly reducing computational overhead and latency, an essential requirement for effective online training. We experimentally demonstrate real-time, end-to-end analog training for both linear and nonlinear classification tasks at gigabaud (GBaud) rates, achieving accuracies of over 90\% and 80\%, respectively. Our analog neuromorphic processor introduces self-learning capabilities that dynamically adjust training parameters, setting the stage for truly autonomous neuromorphic architectures capable of efficient, real-time processing in unpredictable real-world environments. As a result, we showcase adaptive online tracking of dynamically changing input datasets and achieve over 90\% accuracy, alongside robustness to external temperature fluctuations and internal thermal crosstalk---ensuring stable operation under varying environmental changes. This demonstration marks a significant advancement toward self-adaptive, real-time photonic neuromorphic systems, learning bridging the longstanding gap between analog neuromorphic processing and real-world deployment in next-generation artificial intelligence.
}
\keywords{Analog Training, Online Tracking, Adaptive Training, In-Situ Training, Online Learning}

\maketitle

\section{Introduction}\label{sec1}

The recent surge in artificial intelligence (AI) has triggered a paradigm shifts in computation, driving rapid advancements across various scientific and technological fields, including protein folding, natural language processing, computer vision, and autonomous vehicles~\cite{Bywater2010, Zhou2024, Chen2023, 10.1063/1.1144830}. In particular, a wide range of neural network (NN) architectures --- such as deep neural networks, recurrent neural networks, convolutional neural networks and several others --- have been explored, each tailored to specific computational tasks~\cite{LeCun2015, 6639349, Krogh2008}. These architectures primarily rely on numerical computations performed on CPUs and GPUs to train the networks by estimating gradients. Recently, several neuromorphic hardware architectures have been developed to compute gradients directly, including variants of backpropagation, online pruning, and other analog-friendly algorithms~\cite{10.3389/fncom.2017.00024, Xu:24, model_free, Lin2025}. Among all these techniques, backpropagation remains the most widely used algorithm in modern machine learning due to its effectiveness in gradient computation~\cite{Wright2022}.  However, backpropagation requires gradient communication to propagate backward through the network, layer by layer. Furthermore, it often relies heavily on digital platforms for offline training. As a result, training neural networks demands large datasets, intensive matrix multiplications, and gradient computations—leading to high computational costs, bandwidth constraints, and memory bottlenecks ~\cite{Fu2024, Wetzstein2020}. A promising stepping stone towards autonomous training is computer-in-the-loop training, which reduces hardware dependency but incurs significant latency in the training loop~\cite{Filipovich:22, Dalgaty2021, Lin2024, Dohare2024}. However, a fully autonomous online learning system—operating entirely without reliance on an external CPU—represents a key future direction for achieving faster, more energy-efficient, high-performance computing. As a result, various algorithms and architectures have been proposed to accelerate the online training, including physical hardware-aware training, local learning, continuous learning and equilibrium propagation ~\cite{Lillicrap2020, momeni2024trainingphysicalneuralnetworks}. All these approaches demonstrate a transition toward autonomous online learning, known as \emph{in-situ} training, which eliminates the need for an external computer~\cite{BuckleyTaitMcCaughanShastri+2023+833+845}. Implementing in-situ training on physical platforms such as photonics, nanoelectronics, or mechanical systems has the potential to significantly enhance computational efficiency and widen the range of practical applications ~\cite{wang2024opticaltraininglargescaletransformers, doi:10.1126/sciadv.adi9127}. In this work, we demonstrate in-situ, fully end-to-end analog system with an analog algorithm that is agnostic to hardware platforms, potentially including photonics, memristors, analog CMOS chips, and spintronics. Our approach enables neuromorphic hardware systems to be trained as a black box, moving one step closer to truly autonomous online learning. 

In practice, signals natively generated by sensors, spectroscopy, astronomy, high-energy physics, and medical applications need to be processed directly in the analog domain. Our analog neuromorphic photonic processors are uniquely positioned to harness the ultrafast bandwidth and energy efficient, enabling direct processing of analog signals in the optical domain ~\cite{Marpaung2019, Shastri2021, QuerliozDamien2024Psat}. Analog optical computing architectures eliminate the need for high-precision digital multiplications and offer low latency, inherent parallelism, and scalability. As a result, recent advancements in integrated photonic architectures—such as Mach-Zehnder interferometric meshes, microring resonator arrays, metamaterials, and nanostructures—have enabled a wide range of applications, including low-latency signal processing of radio and optical signals ~\cite{Huang2021, Zhang2023, Lederman2023, Zhang2024}, fast matrix multiplications~\cite{Totovic2022, Lin2024}, and image processing~\cite{Xu2021, Feldmann2021}. Furthermore, the adoption of analog photonic hardware for training—incorporating backpropagation, forward propagation, and perturbative approaches—has spurred efforts to develop effective in-situ training methods to fully harness its potential ~\cite{doi:10.1126/science.ade8450, Zheng2023a, Xue2024}. Despite their excellent inference performance, several challenges remain unresolved in state-of-the-art analog photonic architectures ~\cite{Zhang2023, Mourgias-Alexandris2022}. These challenges include limited training precision, the lack of on-the-fly training with dynamic data, the need for frequent analog-to-digital conversions, and the inadaptibility to crosstalk and environmental variations in real-time. In addition, most of the in-situ training algorithms are highly hardware-dependent, unsuitable for high-speed operation, and are difficult to implement across a broad range of neural network platforms. Therefore, unlocking the full potential of photonic neuromorphic hardware requires the development of training techniques that accommodate diverse device models while remaining robust to device-to-device variations and noise.

\begin{figure}[pthb]
    \centering
    \includegraphics[width=1\linewidth]{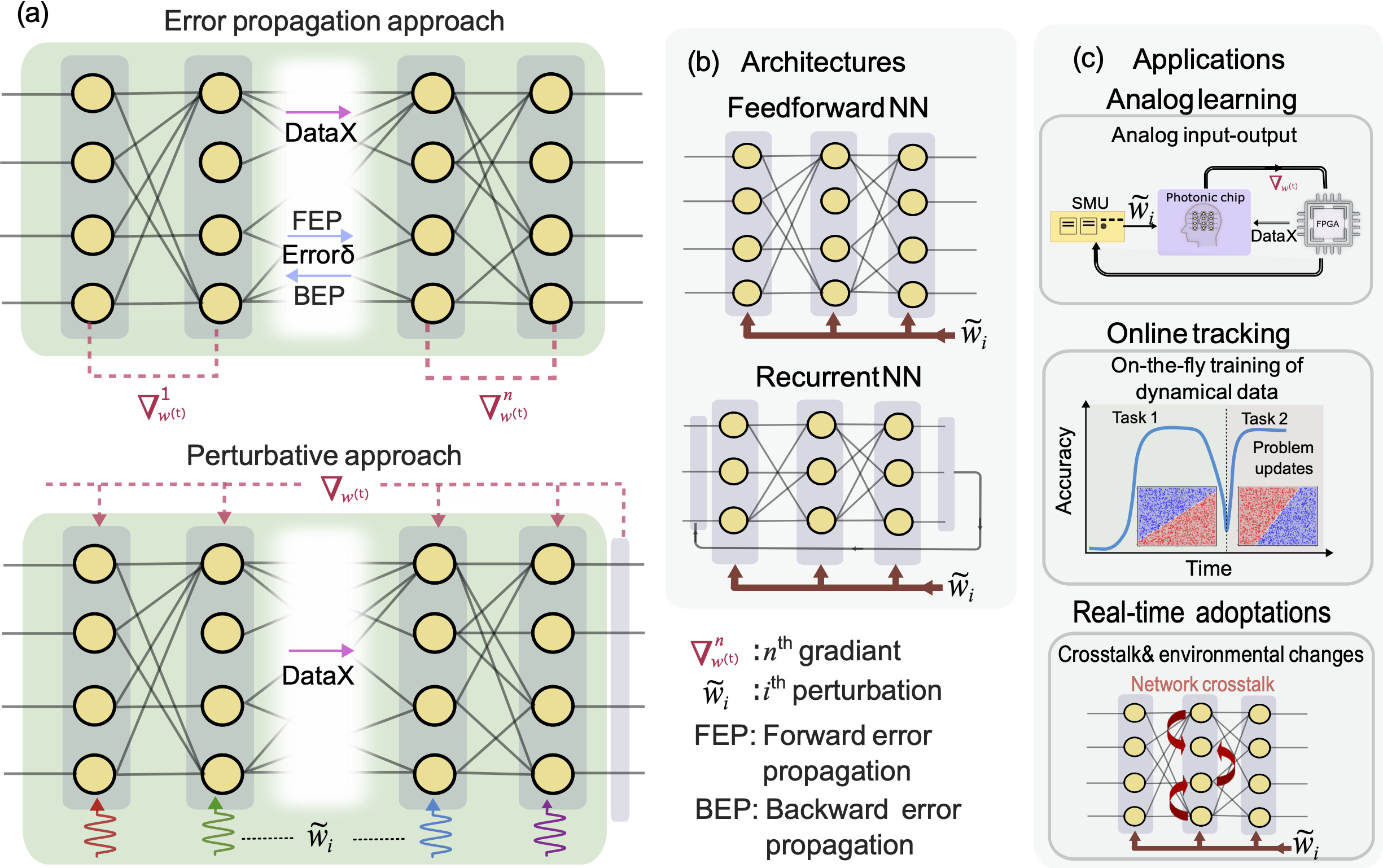}
    \caption{{\bf Artificial intelligence training approaches and the analog end-to-end adaptive training system: }a) Widely adopted training methods implemented on analog learning hardware platforms include error propagation and perturbative approaches. b) Demonstrated by the multiplexed gradient descent algorithm~\cite{10.1063/5.0157645}, perturbative approaches can be applied to a wide range of machine learning models, including feedforward neural networks and recurrent neural networks. c) These models give rise to a variety of real-life applications, enabling hardware-agnostic analog learning directly using analog signals. This further enables our analog hardware to use its high bandwidth and low latency in online tracking tasks, and creates the flexibility to adapt to various system- and platform-level parameter changes such as crosstalk.}
    \label{fig:1}
\end{figure}

Here, we experimentally demonstrate in-situ, fully end-to-end analog and on-the-fly training on an integrated photonic platform using the multiplexed gradient descent (MGD) algorithm~\cite{10.1063/5.0157645}. In this approach, the gradient of the signal is directly measured at the output, eliminating the need for digital gradient calculations using matrix multiplications. The end-to-end analog architecture physically measures the gradient and locally updates weights, so the entire training pipeline remains within the analog domain. Gradient measurement is achieved by injecting small perturbative signals into the network parameters and monitoring variations in the loss function based on device dynamics~\cite{10.1063/5.0157645}. The flexibility and hardware-compatible nature of the algorithm enable the implementation of on-chip photonic analog end-to-end adaptive training on a foundry-manufactured and CMOS-compatible silicon photonic integrated circuit (PIC). Our architecture efficiently classifies both linear and nonlinear tasks, achieving accuracy rates exceeding 90\% and 80\%, respectively. Additionally, our adaptive system can dynamically track environmental changes, such as temperature variations and internal thermal crosstalk, by adjusting training parameters in real time without external monitoring or downtime. We further demonstrate adaptive online tracking at a Giga inference-per-second (GBaud rate), achieving over 90\% accuracy with rapid training convergence. This high-speed, low-latency in-situ, on-the-fly training enables real-time adaptation and online tracking, overcoming the limitations typically encountered in other photonic architectures. In addition, we performed a simulation using a $4-30-3$ deep network of the Yin-Yang dataset, consisting of a total of 210 weights~\cite{kriener2022yinyangdataset}, achieving an accuracy of over 97\%, demonstrating the scalability of the proposed method at larger scales. Our demonstration is hardware-agnostic, robust against system noise and fabrication variations, and does not require sophisticated calibration routines~\cite{9851450}. These characteristics mark a significant advancement toward self-adaptive, real-time photonic neuromorphic systems capable of efficient online learning in complex, real-world scenarios. In addition, our proposed analog end-to-end training system lays the foundation for scalable, high-speed, and energy-efficient photonic machine learning platforms, bridging the longstanding gap between analog neuromorphic processing and real-world deployment in next-generation artificial intelligence.

\section{Hardware implementation}\label{sec2}

\begin{figure}[pthb]
    \centering
    \includegraphics[width=1.0\linewidth]{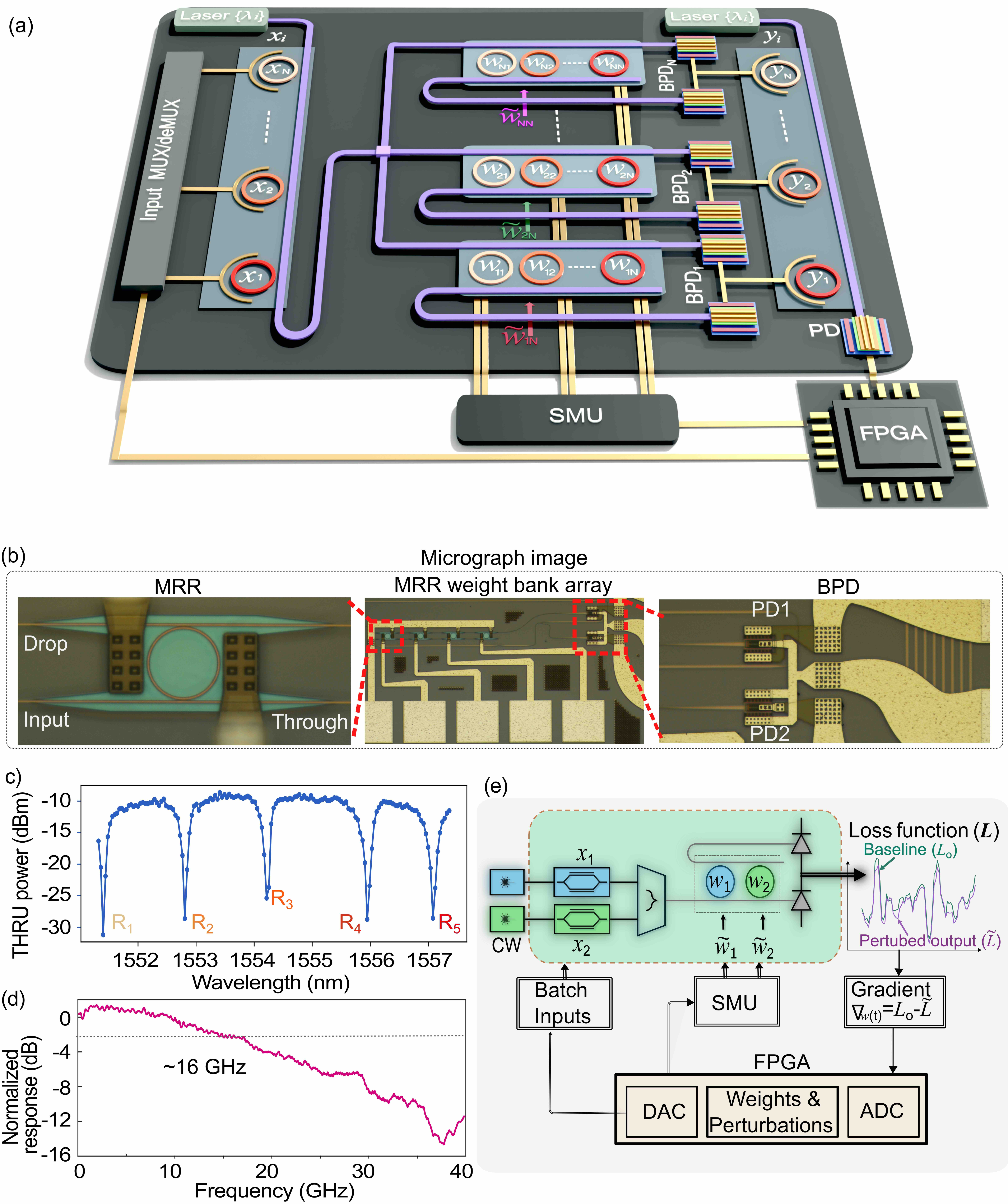}
    \caption{{\bf Schematics for experiment setup and device characterizations: }a) Our vision for a wavelength-muliplexed integrated neuromorphic photonic hardware that uses microring resonator (MRR) weight bank and on-chip balanced photodetectors (BPDs). The inputs $x_i$ are amplitude-encoded on respective wavelength channels using $i^th$ microring modulators, all of which are broadcast to all neurons in the same layer. Each weight bank implements all the weights $w_{NN}$ for the particular neuron plus the corresponding perturbations $\widetilde{w}_{NN}$, and the balanced photodetector BPD$_N$ signals are sent to N microring modulators. The biasing of the microring modulator determine the types of the nonlinear activation function and produces output $y_i$ for that neuron. b) Micrographs of the electrically-packaged integrated photonic chip, and the device-under-test, with labels for the electrical connections for the weights and photodetector bias. c) Spectrum of the MRR weight bank. d) Bandwidth measurement of the integrated BPDs. e) The experimental setup used in this paper: a single-layer, 2-input and 2-weight system consists of two optical input channels, $x_1$ and $x_2$, each modulated by a DAC on the RFSoC FPGA via a Mach-Zehnder modulator (MZM). A WDM arrayed-waveguide grating (WDM AWG) multiplexes both channels into a single waveguide, which is fed to the add-drop MRR weight bank integrated on chip. The MRRs are controlled using two SMU channels, which communicate with the RFSoC FPGA to actuate the weights $w_1$, and $w_2$, as well as perturbations $\widetilde{w}_1$ and $\widetilde{w}_2$. The weight bank optical output is detected through a pair of integrated BPDs, and the output signals $L_0$ and $\widetilde{L}$ are measured with an ADC channel on the RFSoC FPGA.}
    \label{fig:2}
\end{figure}

The schematic illustration of the analog end-to-end training is implemented on a wavelength division multiplexing (WDM) platform within a fully integrated on-chip system is shown in Fig.~\ref{fig:2}(a). The chip consists of tunable filters, each independently tuned to a specific wavelength channel. The photonic neural network illustrated in Figure 2(a) comprises an input modulator for encoding optical signals, a weight bank array, balanced photodetectors (PDs), and PN-junction-based modulators for the nonlinear activation function. Both the input-output modulators and weight bank arrays are implemented using an array of compact circular waveguides known as microring resonator (MRR). The micrograph of the fabricated WDM photonic neural network, including the microring resonator weight bank and balanced photodetectors, is shown in Figure 2(b).  The drop responses of the microring weight bank array and the measured bandwidth of the PD are shown in Fig.~\ref{fig:2} (c) and (d), respectively. The input signals are amplitude-encoded across \textit{n} wavelengths in an $m \times n$ network array. The microring weights are controlled by source measurement units (SMUs), which set the voltage for each MRR. These voltages actuate MRR resonances via in-resonator photoconductive heaters (IRPHs) using the thermo-optic effect~\cite{Jayatilleka:15}. The weighted input values are summed at the balanced photodetector, resulting in efficient matrix multiplication. This process converts the analog output signals independently into the electrical domain using \textit{m} balanced photodetectors. The incoherent WDM broadcast and weight architecture eliminates the need for precise control of phase shifters, mitigating interference and instability commonly encountered in architectures based on coherent matrix multiplications. To enable fast data encoding, weight updates, and gradient detection, we employ digital-to-analog converters (DACs) and analog-to-digital converters (ADCs) integrated into an RFSoC FPGA. The DAC operates at a baud rate of 1 GBaud with 8 samples per symbol, while the ADC samples at 4 GS/s. Further details of the experimental setup are provided in the Methods section. Figure ~\ref{fig:2}(e) presents a schematic of the implemented experimental setup.

The instantaneous loss function variations, $\widetilde{w}(t)$, consists of a perturbation-free baseline loss function $L_0$, and a perturbed loss function $\widetilde{L}_i(t)$ with a variation component $\Delta L_i(t)$, generated by the weight perturbation, $\widetilde{w}_i(t)$, at each neuron. Although the weight perturbations can take any pairwise-uncorrelated time-varying function, we use a randomly generated binary perturbation for simplicity. The perturbation $\widetilde{w}_i(t)$ follows a random binary sequence of $(-1,1)$, with an amplitude of $\Delta w_i$. Each weight is implemented as a voltage signal applied to the MRRs in the weight bank, with perturbations introduced as the sum of both the weight and its perturbation. The binary choices in the random sequence are unique for each batch during every training epoch. As training progresses, this random sequence generates a time-varying perturbation signal, causing corresponding variations in the loss function measurements. The loss function is computed using the network output, $y(t)$, and the training target $\hat{y}(t)$. The weight update $w_i$ incorporates the instantaneous measurements of the variation component of the loss function and its corresponding weight perturbation. More details on the derivation of the MGD algorithm for directly measuring the gradient are provided in the \textbf{Supplementary Section 1.1}. Our chip performs analog gradient approximations with perturbative training techniques~\cite{Dembo1990a}, eliminating the need for conventional large-scale matrix multiplications typically used for gradient calculations in digital learning algorithms.

\section{Results}
\subsection{Analog end-to-end training}\label{subsec1}

\begin{figure}[pthb]
    \centering
    \includegraphics[width=1\linewidth]{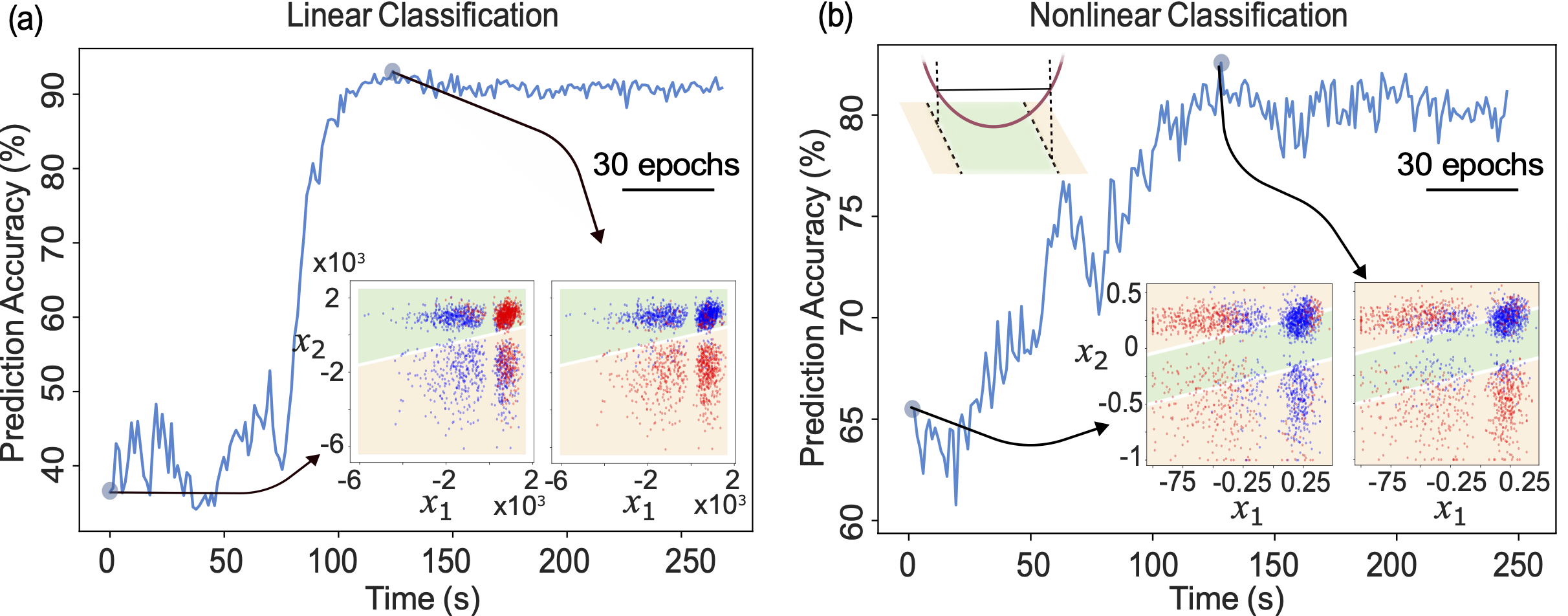}
    \caption{{\bf Analog end-to-end training results: }experimentally measured training accuracy values in both linear and nonlinear classification problems, plus the system stability validation results. a) Linear classification of two sets of data trained over 200 epochs, and both the initial and the optimal accuracy results are shown in the inset scatter plots. The classification rule is defined by an arbitrary line ($0.0137\cdot x_1-1\cdot x_2=0$). The background colors indicate the correct classification results, and the color of the data points represents the results produced by the network. b) Nonlinear classification with a quadratic separation of two sets of data, also trained over 200 epochs. The nonlinear classification rule is defined by an arbitrary parameter ($(0\cdot x_1+1\cdot x_2)^2-0.05=0$), which can be visualized as the projection of a parabola onto a 2D plane as shown in the inset. The classification results are shown in the two scatter plot insets, where the background colors indicate the correct classification results, and the color of the data points represents the results produced by the network.}
    \label{fig:3}
\end{figure}

First, we demonstrate our analog end-to-end adaptive training system by solving a linear classification problem, where the classification rule is defined by an arbitrary line ($a\cdot x_1-b\cdot x_2=0$), where $a$ and $b$ are real numbers. In the experiment, all the network parameters are initialized to random values, and the gradient is measured by comparing baseline and perturbed loss function values --- without and with perturbations to the network parameters, respectively. These parameters are updated after each batch, and training continues until convergence is reached. As shown in Fig~\ref{fig:3}(a), our system achieves convergence in under 100 seconds, which is comparable to the calibration time in other calibration-based approaches. We train the system for 200 epochs, achieving a prediction accuracy exceeding 90\% with no significant fluctuations. As a result, the standard deviations of the prediction accuracy and loss function are $0.259$ and $0.00312$, respectively.

Furthermore, we have trained the system for a nonlinear classification task using the same approach. Figure~\ref{fig:3}(b) shows a demonstration of a nonlinear classification problem that introduces an arbitrary quadratic nonlinearity ($(a\cdot x_1+b\cdot x_2)^2-0.05=0$) at the photodetector output. Unlike linear classification, the output of the quadratic function exists in the third dimension, where it intersects the original 2D plane to form a band that defines class separation, as illustrated in Fig.~\ref{fig:3}(b) insets. For the experiment, we follow the same procedure used in the linear classification case, with the addition of the quadratic nonlinearity programmed on the RFSoC FPGA. As shown in Fig.~\ref{fig:3}(b), our system reaches convergence within 100 training epochs, achieving a maximum accuracy above 80\%. After convergence, the prediction accuracy remains stable, with a standard deviation of 1.06\%.

\subsection{Online tracking}\label{subsec2}

Online tracking is a time-sensitive task; consequently, real-time forecasting, anomaly detection, object tracking, and adaptive control systems require continuous monitoring of dynamically evolving data~\cite{Yang2024, Blaser2000, Lale2024}. Adapting to such data necessitates rapid and continuous parameter updates, dynamic learning rules, and operation across varying conditions—all while ensuring timely decision-making. Implementing a hardware-agnostic algorithm capable of adapting to environmental variations typically involves updating samples, modifying learning rules, and/or adjusting operating conditions. Existing photonic computing hardware struggles to perform fully online real-time tracking due to (1) the need to simultaneously execute training and inference, and (2) the requirement to achieve faster convergence and parameter update. Therefore, to successfully implement online tracking, the hardware must update network weights on-the-fly as soon as new data arrives. Here, we address the online tracking problem by enabling simultaneous training and inference with fast convergence, while the learning rule for classification continuously adapts. The dataset consists of 2,000 samples uniformly distributed in a 2D space, with the correct separation between two classes evolving over time as reflected by updates in the learning rule signal.

\begin{figure}[pthb]
    \centering
    \includegraphics[width=1\linewidth]{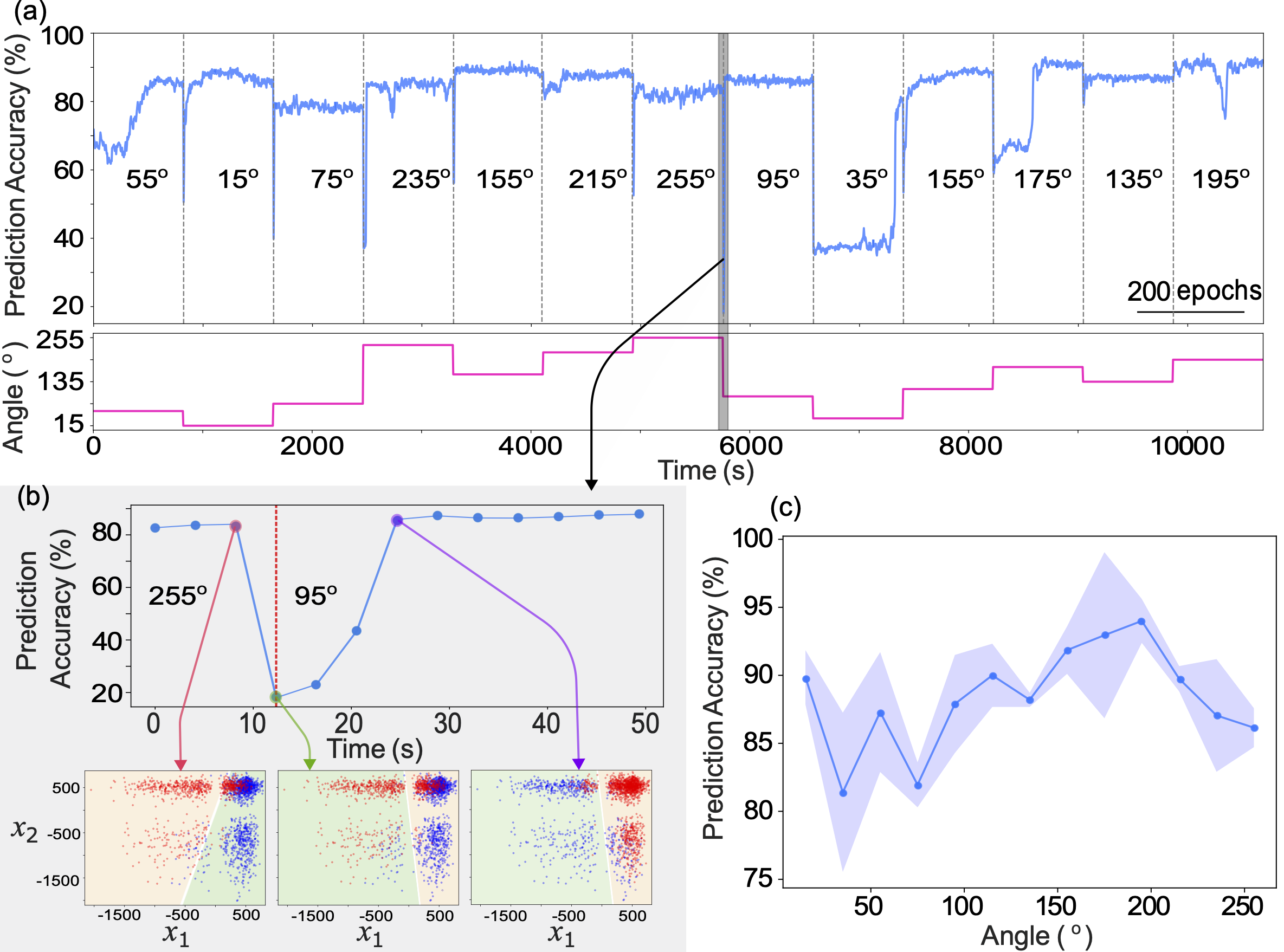}
    \caption{{\bf Online tracking results: }the separation rule rotates through a sequence of three different angles calculated with respect to the x-axis, and we demonstrate our system successfully adapts to the logical changes. a) Training accuracy values measured experimentally at thirteen unique angles ranging from $15^\circ$ to $255^\circ$ with a step size of $20^\circ$, updated after every 200 epochs. The thirteen different angles are randomly shuffled, and are plotted against time in the subplot here. b) Zoomed-in plot of the epochs near the transition from $255^\circ$ to $95^\circ$, together with three scatter plots showing the system output before, at, and after the angle change. c) Best prediction accuracy achieved at every angle, plotted with the corresponding standard deviation recorded.}
    \label{fig:4}
\end{figure}

For the online tracking task, we varied the classification criteria every 800 seconds — equivalent to 200 training epochs — without providing any additional information to our analog system during training. Initially, the separation between the two classes was set at a $55^\circ$ angle relative to the horizontal axis. Subsequently, the separation between the two classes was rotated through 13 randomly shuffled configurations, with angles ranging from $15^\circ$ to $255^\circ$ in $20^\circ$ increments. The sequence of shuffled angles is illustrated in the subplot of Fig.~\ref{fig:4}(a). At the beginning of each angle change, there is an initial drop in prediction accuracy, as seen in Fig~\ref{fig:4}(a). However, the system quickly recovers to the new problem, achieving a prediction accuracy of approximately $>$85\% within a few epochs ($<$10) for most angles. We also measure the training time in real-time, and this performance corresponds to a recovery time of approximately 20 seconds whenever the learning rule is updated. More specifically, a zoomed-in plot of the 13 epochs around the transition from $255^\circ$ to $95^\circ$, shown in Fig~\ref{fig:4}(b), validates that our system restores its former performance within just 4 epochs, corresponding to a training time within 20 seconds. Figure~\ref{fig:4}(c) shows the best prediction accuracy achieved at every angle, plotted with the corresponding standard deviation recorded. However, we occasionally observe a momentary decrease in prediction accuracy after the system has reached convergence,  primarily as a byproduct of the perturbations used in our analog training method. While such fluctuations can be mitigated through fine-tuning hyperparameters—such as perturbation amplitude and learning rate—they cannot be entirely eliminated due to the inherent nature of the perturbation-based approach. Notably, such brief moments of instability only occur sporadically and have a negligible impact on overall system performance.

\subsection{Real-time adaptation}\label{subsec3}

\begin{figure}[pthb]
    \centering
    \includegraphics[width=1\linewidth]{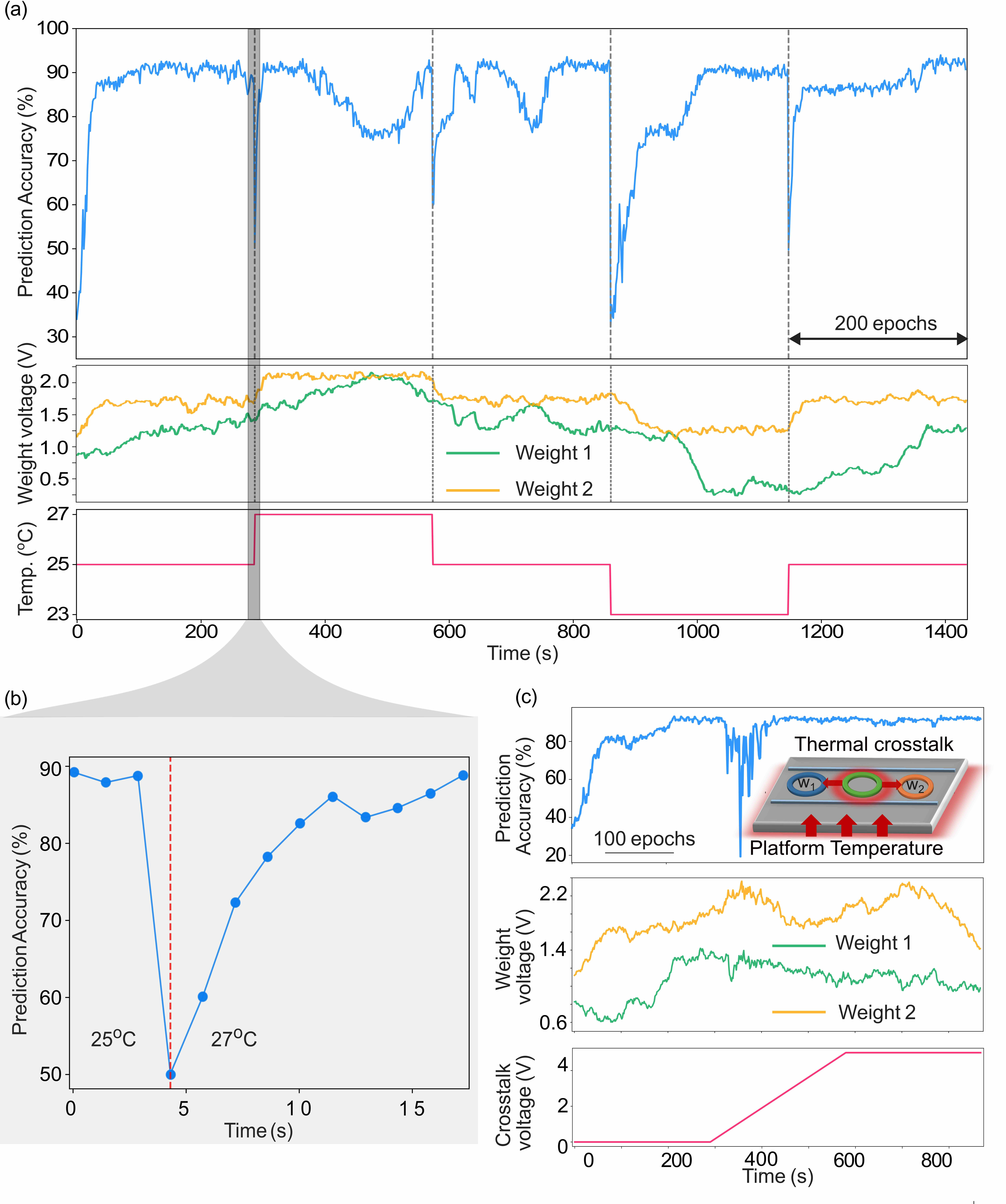}
    \caption{{\bf Real-time adaptation results:} we demonstrate the excellent stability of our system against two major environmental variations. a) we apply temperature changes to the test platform by varying the target temperature set by the temperature controller every 200 training epochs. The temperature starts at $25^\circ\rm C$, and goes to $27^\circ\rm C$, $25^\circ\rm C$, $23^\circ\rm C$, and $25^\circ\rm C$ subsequently. We record training accuracy after each epoch and provide a zoomed-in plot of the epochs near the first temperature transition from $25^\circ\rm C$ to $27^\circ\rm C$. At $25^\circ\rm C$, the actuated weight voltage range of $[0, 2]$ V corresponds to a measured optical weight range of $[-1, 1]$. b) we introduce on-chip thermal crosstalk by injecting a steadily increasing bias voltage to the channels neighboring the ones under test. Starting from the $200^{\rm th}$ epoch, the bias voltages to the neighboring two channels are increased from 0 V to 5 V over a period of 200 epochs, with a step size of 25 mV per epoch. The thermal crosstalk persists after the bias voltages reach the maximum value for the rest of the trial. c) Weight voltage values recorded throughout the thermal crosstalk demonstration, together with the bias voltage supplied to the neighboring channels to generate thermal crosstalk.}
    \label{fig:6}
\end{figure}

Analog computing hardware is typically sensitive to both external and internal variations—such as environmental temperature changes and thermal crosstalk—which can significantly affect system performance~\cite{Jayatilleka:15, Zhang:22, Zhang2023}. For example, environmental temperature fluctuations and thermal crosstalk between neighboring heat-actuated weights degrade reliability. While some performance gains have been achieved using error correction and hybrid analog-digital architectures, our analog system continuously adapts and readjusts in real time, maintaining stable performance during simultaneous training and inference. As a result, the photonic analog system offers fully integrated self-learning and self-correction capabilities, enabling dynamic adaptation to both external non-idealities (e.g., platform temperature changes) and on-chip non-idealities (e.g., thermal crosstalk) in real time, using the MGD algorithm for simultaneous inference and training.

Our photonic chip is placed on a temperature-controlled stage that maintains the operating temperature within $\pm$0.25°C of the target set value. Typically, a temperature variation of just 2°C can result in up to a 50\% change in the effective weights. As a proof of concept, we first trained the system at $25^\circ\rm C$ for 200 epochs. At this temperature, optical weight values of 1 and -1 correspond to actuation voltages of 0 V and 2 V, respectively. Within this range, the mapping between the applied voltage and the resulting optical weight is approximately linear. We then increase the platform temperature to $27^\circ\rm C$ and continue training for another 200 epochs. Subsequently, we adjust the temperature to $25^\circ\rm C$, then to $23^\circ\rm C$, and back to $25^\circ\rm C$, changing the temperature every 200 epochs. The training accuracy per epoch is shown in Fig.~\ref{fig:6}(a), demonstrating that our system can accommodate platform temperature variations up to $\Delta T (\pm2^\circ\rm C)$ and quickly recover its performance after each temperature change, as confirmed in the zoomed-in plot in Fig.~\ref{fig:6}(a). 

To better explain the effect of a temperature change $\Delta T$ on the weights, we introduce the concept for the rate of equivalent weight change which can be calculated as $\text{\em rate of equivalent weight change} = {\Delta w_o}/{t_T}$, where $\Delta w_o$ is the equivalent percentage change in the optimal weight value as a result of the temperature change, and $t_T$ is the amount of time between temperature change measured in minutes. Because the target problem is constant, the optimal weights should have the same numerical values despite the changes to the physical parameters. Here, the average time between temperature change is approximately 4.8 minutes, and we find that the rate of equivalent weight change to be around $23.7\%\cdot{\rm min}^{-1}$ for weight 1 and around $5.48\%\cdot{\rm min}^{-1}$ for weight 2. This indicates that our system is capable of accommodating environmental temperature changes that translate to a weight change of more than 20\% per minute. This rate could be adapted by varying the learning rate. To assess system stability against thermal crosstalk, we steadily increase the bias voltages two neighboring channels from 0 V to 5 V over 200 epochs, as shown in Fig.~\ref{fig:6}(b). This adjustments occur after the initial 200 epochs, with the bias voltage increasing by 25 mV per epoch. Following this, we introduce sustained thermal crosstalk by maintaining a 5 V bias voltage on both neighboring channels for the final 200 epochs. As shown in Fig.~\ref{fig:6}(b), brief moments of instability occur as the bias voltages ramp up; however, our system quickly recovers and remains resilient against the effects of thermal crosstalk.

\section{Discussion}

Our system demonstrates in situ, on-the-fly training capabilities while addressing key challenges in neuromorphic hardware, such as stability in online tracking and real-time adaptation. We validated the effectiveness of the MGD method on a foundry-manufactured silicon photonic integrated circuit. Additionally, we evaluated system performance metrics, including scalability, latency, convergence time, and energy efficiency. The estimated key metrics and performance comparison with state-of-the-art architectures are summarized in Table~\ref{tab1}. 

\begin{figure}[pthb]
    \centering
    \includegraphics[width=0.9\linewidth]{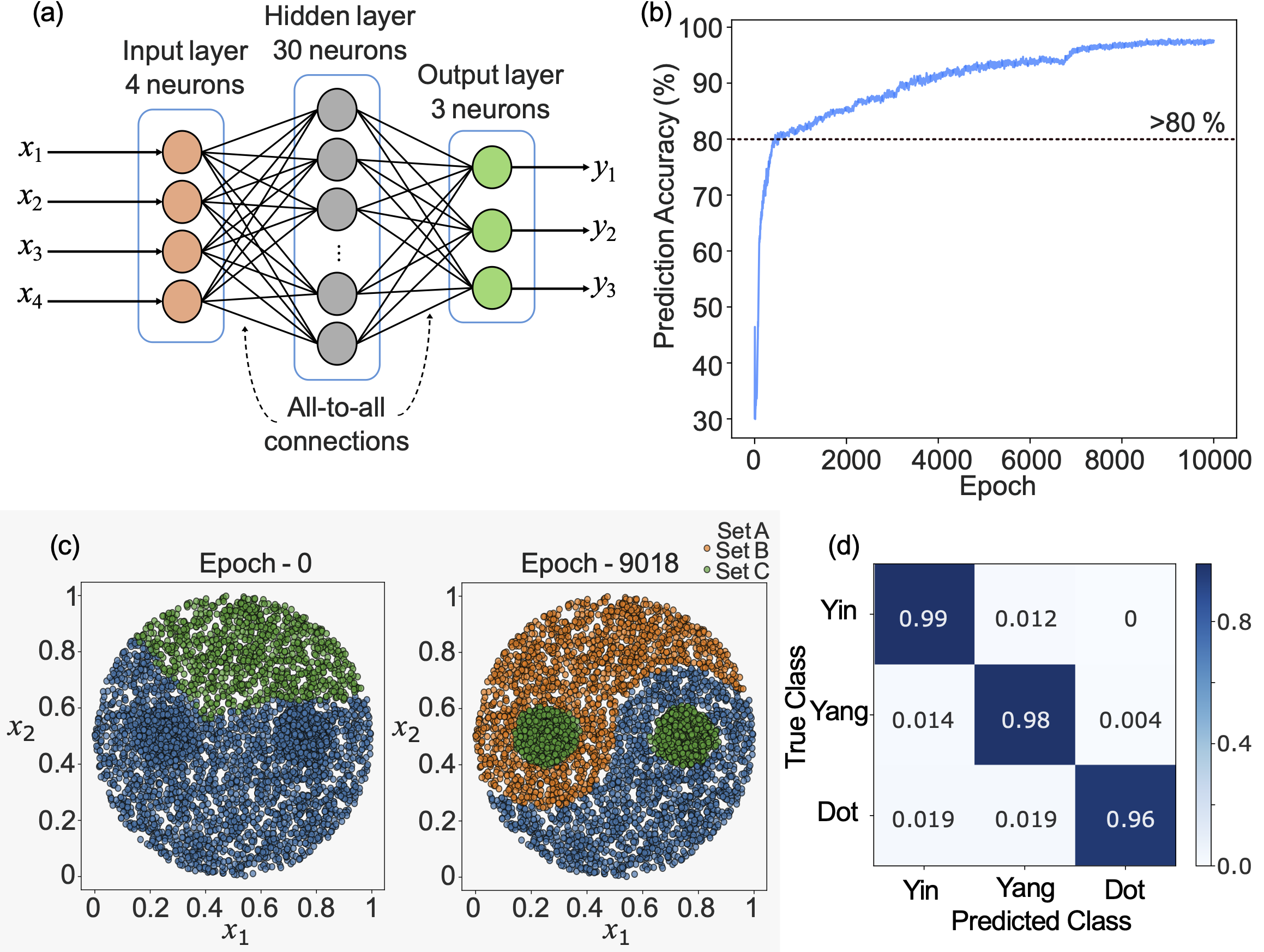}
    \caption{{\bf Yin-yang classification with multi-layer networks: }we simulated our analog end-to-end adaptive training system with a feedforward network containing one input layer with 4 neurons, one hidden layer with 30 neurons, and one output layer with 3 neurons, all fully connected as shown in a). All neurons in both the hidden layer and the output layer feature a ReLU nonlinearity, and include a trainable bias for each node. b) Training accuracy results obtained from our simulation over 10,000 epochs, reaching a maximum accuracy of 97.9\%. c) Snapshots of the classification outputs at the beginning of the training and at the epoch with the best accuracy value. d) Confusion matrix at the end of the training simulation.}
    \label{fig:5}
\end{figure}

The MGD algorithm has shown excellent scalability in digital electronic hardware~\cite{10.1063/5.0258271}. Extending this concept, we simulate the MGD training of a multi-layer network implemented on an analog end-to-end adaptive photonic training system. The network is trained to classify the Yin-Yang dataset~\cite{kriener2022yinyangdataset} comprising an input layer of four nodes, a hidden layer of 30 nodes, and an output layer of three nodes, totaling 37 neurons in a fully connected configuration, as shown in Fig.~\ref{fig:5}(a). In the hardware architecture, each node would be mapped to four MRRs, resulting in a total of 210 MRRs. Furthermore, we incorporated a rectified linear unit (ReLU) nonlinearity in the simulation to enhance the training efficiency. This nonlinearity can be realized by appropriately biasing the microring modulator neurons~\cite{PhysRevApplied.11.064043}. Our large-scale multi-layer network simulation follows the procedure outlined in Algorithm 1, where weight and bias perturbations are simultaneously introduced in each batch. The training progress is depicted in Fig.~\ref{fig:5}(b), where the system achieves over 80\% accuracy around 1,000 epochs and reaches 97.9\% accuracy after 8,000 epochs. The classification results and the corresponding confusion matrix at the end of training are shown in Fig.~\ref{fig:5}(c) and Fig.~\ref{fig:5}(d), respectively. For comparison, a CMOS-based implementation trained using the ADAM optimizer in PyTorch achieved $97.6\pm1.5\%$ accuracy, demonstrating parity with our photonic approach. These findings validate the scalability and effectiveness of perturbative training techniques for photonic circuits in larger-scale networks.

Furthermore, we analyze the system latency of our analog photonic platform. The time delay between the input vector and the output classification is 11.7 ps. When accounting for the response time of high-speed ADCs, DACs, and photodetectors, the total end-to-end classification latency remains low at 74.2 ps (see the Supplementary Section). Convergence time is another critical metric for training. We experimentally measured convergence times across different tasks under a fixed operating condition. For classification tasks, the total convergence time is $\sim$15s. The frequency and abrupt variations in internal parameters significantly influence the time required to exceed 80\% accuracy. The system takes longer to converge for slower processes, such as temperature variations. In the current configuration, the N$\times$N network has N$^{2}$ control parameters for matrix multiplication, with weight elements currently being temperature-controlled, contributing to the observed convergence time. However, our photonic chip also supports carrier-depletion PN junction for electro-optic weight tuning, which provides a typical bandwidth of 17 GHz ~\cite{hagan2020high}. This enables rapid weight adjustments, reducing convergence times by a factor of 1,000 from microseconds to milliseconds depending on the application.

\begin{table}[pthb]
\caption{Comparisons with other contemporary works.}\label{tab1}%
\begin{tabular}{p{0.25\linewidth} | p{0.08\linewidth} | p{0.1\linewidth} | p{0.2\linewidth} | p{0.09\linewidth} | p{0.1\linewidth}}
\toprule
  & Analog inference latency & Analog training latency & Energy efficiency & Online tracking~\footnotemark[1] & Real-time adaptation~\footnotemark[2] \\
\hline
\textbf{This work}  & \textbf{11.7 ps} & \textbf{22 ms} & \textbf{3,000 GOPS/W} & \textbf{Yes} & \textbf{Yes} \\
Momeni, A. {\em et al.}~\cite{doi:10.1126/science.adi8474} & 192 ps & Not given & 0.12 GOPS/W & No & No \\
Bandyopadhyay, S. {\em et al.}~\cite{Bandyopadhyay2024} & 410 ps & Not given & 13 GOPS/W & No & No \\
Pai, S {\em et al.}~\cite{doi:10.1126/science.ade8450} & 1 ns & Offline training & 5,000 GOPS/W & No & No \\
Xue, Z {\em et al.}~\cite{Xue2024} & Not given & 64 ms & $5.4\times10^{9}$ GOPS/W & No & No \\
Ashtiani, F. {\em et al.}~\cite{Ashtiani2022} & 570 ps & Digital training & 71.4 GOPS/W & No & No \\
\botrule
\end{tabular}
\footnotetext[1]{Online tracking of updates to the training problem or learning rule.}
\footnotetext[2]{Real-time adaptation to external and internal parameter changes, including environmental temperature variations and neighboring channel thermal crosstalk.}
\end{table}

Our work underscores the energy efficiency of the monolithically integrated photonic chip. The system has a total power consumption of 27.4~mW--- including photonics, drivers, and readout electronics, resulting in an energy efficiency of 853 fJ/Op, equivalent to 1.2~TOPS/W at a 16~GBaud operational speed. The photodetector bandwidth of 16 GHz is currently the limiting factor for higher inference rates in our setup. See the Supplementary section for a detailed breakdown of the power budget and energy efficiency of individual components. For the 4$\times$30$\times$3 neural network architecture (see Fig.\ref{fig:5}(a)), each neuron is implemented with four MRRs, performing 243 operations (5$\times$30$+$31$\times$3) per inference cycle, yielding an energy efficiency of 3.4 fJ/Ops, corresponding to 291 TOPS/W. A comparison of on-chip and free-space optical computing architectures is presented in Table~\ref{tab1}. Moreover, we highlight that integrating phase-change materials (PCMs) in our architecture could push energy efficiency to the PetaOPs/W range while significantly enhancing information density. This capability paves the way for ultrafast, power-efficient, and low-cost neuromorphic computing platforms, addressing both scalability and real-time adaptability in photonic machine learning.

\section{Conclusion}\label{sec5}

We have proposed and experimentally demonstrated an analog end-to-end adaptive neuromorphic hardware system that physically measures and locally updates the gradient of the dynamical system directly in the optical domain. Unlike conventional digital implementations, which are constrained by the limitations of digital matrix multiplications, our platform achieves fully analog gradient measurement through multiplexing and perturbative updates at the output. This approach effectively bypasses digital precision bottlenecks and high-accuracy inference while maintaining robust real-time adaptability. Our demonstration addresses critical, longstanding challenges that have limited the practicality of neuromorphic hardware architectures, particularly the inability to perform on-the-fly training with dynamic data and the lack of real-time adaptation to crosstalk and environmental fluctuations. Implemented on an integrated photonic platform, our system achieves favorable performance in online tracking tasks, reaching 86.5\% accuracy while dynamically adapting to learning rules. Notably, our system ensures stability even when training continues past convergence, eliminating the need for manual calibration or active stabilization. This intrinsic stability allows our photonic processor to self-adapt to changing conditions with fast convergence, effectively mitigating the impact of system noise. We further evaluated our system’s performance in terms of scalability, latency, and energy efficiency. Our photonic system achieves an ultra-low latency of 11.7 ps and an energy efficiency of 853~fJ/Op, highlighting its potential for large-scale neuromorphic networks. Our platform’s fully analog processing significantly reduces the complexity associated with training neuromorphic hardware, paving the way for scalable implementations in integrated photonics. This work represents a significant advancement toward practical, energy-efficient, high-speed, and noise-resilient analog neuromorphic hardware. The demonstrated online adaptive training approach is versatile and broadly applicable across a spectrum of machine learning platforms. Crucially, it lays the groundwork for real-time learning in analog hardware, with transformative implications for applications in sensing, autonomous driving, telecommunications, and beyond.

\section{Methods}\label{sec3}

\subsection{Device fabrication}

The device-under-test here is fabricated on a silicon-on-insulator (SOI) wafer with a silicon thickness of 220 nm and a buried oxide thickness of 2 ${\rm \mu m}$. The bus waveguides have a width of 500 nm. The MRRs have radii of $8.0\;{\rm \mu m}$, $8.01213\;{\rm \mu m}$, $8.02426\;{\rm \mu m}$, $8.03639\;{\rm \mu m}$, $8.04852\;{\rm \mu m}$. The gap between the ring and the bus waveguide is 200 nm, yielding a Q factor of $\sim 6000$, and the free spectral range is around 12 nm for an MRR with 8 ${\rm \mu m}$ radius. The MRRs have in-resonantor photoconductive heaters~\cite{Jayatilleka:15} that can actuate the weight by thermally tuning the MRR resonance. To implement the N-doped heater, each MRR consists of a circular waveguide etched to a 90 nm thick pedestal that hosts the phosphorous dopants. A 10 ${\rm \mu m}$ wide N doping section is patterned to follow the MRR, outside of which heavy N++ doping is used to make ohmic contacts. The phosphorous dopant concentrations are N: $5\times10^{17}\;{\rm cm^{-3}}$ and N++: $5\times10^{20}\;{\rm cm^{-3}}$. Metal vias and traces are deposited to connect the heater contacts of the MRR weight bank to electrical metal pads. An integrated balanced photodetector sums up the difference between the drop and through ports without the need for a differential amplifier, giving weighted addition of all the input signals. On-chip detection also removes the need to couple the light off-chip, reducing round-trip coupling loss by half. The BPD implements two germanium-based photodetectors connected in series with the n-contact of one PD connected to the p-contact of the second PD~\cite{Hai:13} and has been benchmarked to reach 16 GHz bandwidth.

\subsection{Experiments}\label{subsec5}

In our experimental setup we use an RFSoC FPGA (Xilinx RFSoC $4\times2$) to modulate the inputs and measure the outputs from our integrated photonic weight bank, whose weights are controlled using 4 SMU channels (Keithley 2606B). RF amplifiers (Mini-Circuits ZX60-14012L-S+) are also used to boost the output photocurrent, and we also use bias-tees to split off the DC bias which is measured with another SMU channel. 

The dataset used in both online tracking and stability validation contains 20,000 samples, which are divided into batches each consisting of 100 samples. Every sample has two parameters representing its coordinates in the 2D domain, and each parameter is modulated by a DAC on the RFSoC FPGA.As shown in Fig~\ref{fig:2}(b), every training epoch iterates through all the batches in the dataset, and every batch starts by instantiating all input samples within a batch using the two DACs on the RFSoC FPGA. First we measure the output signal which gives us the baseline loss function values. After perturbing the weights, we measure again the loss function with variations, and the difference between the perturbed and unperturbed loss function corresponds to the gradient.

\backmatter

\bmhead{Acknowledgements}

The U.S. Government is authorized to reproduce and distribute reprints for governmental purposes notwithstanding any copyright annotation thereon. This research was funded by Queen's University (https://ror.org/02y72wh86), University of Pittsburgh (https://ror.org/01an3r305), and NIST (https://ror.org/05xpvk416).

Thanks to Nathanael Eddy for his assistance taking the micrographs using a modified Nikon-E microscope (Centre for Nanophotonics, Department of Physics, Engineering Physics, and Astronomy, Queen's University).

\bibliography{sn-bibliography}

\end{document}